\documentclass[12pt]{article}
\usepackage{graphics}
\usepackage{amssymb,epsfig} %amsfonts 

\author{{\sc K.-H. Rehren} \\
Institut f\"ur Theoretische Physik, Universit\"at G\"ottingen
(Germany)}
\title{\vskip-16mm\bf Algebraic Holography}
\def\frac#1#2{{#1\over#2}} \def\a{\alpha} \def\inv{^{-1}}
\def\ZZ{\mathbb{Z}}  \def\RR{\mathbb{R}} 
\def\ads{AdS_{1,s}} \def\cm{C\!M_{1,s-1}}

\def\ccdot{\!\cdot\!} \def\qed{\hspace*{\fill}$\square$} 
 \def\wt{\widetilde W} \def\pw{W}
\begin{document}
\renewcommand{\today}{}

\maketitle %\vskip5mm
{\bf Abstract:} {\small A rigorous (and simple) proof is given
that there is a one-to-one correspondence between causal
anti-deSitter covariant quantum field theories on anti-deSitter space
and causal conformally covariant quantum field theories on its conformal
boundary. The correspondence is given by the explicit identification
of observables localized in wedge regions in anti-deSitter space and 
observables localized in double-cone regions in its boundary. It takes
vacuum states into vacuum states, and
positive-energy representations into positive-energy representations.}

\section{Introduction and results}

The conjectured correspondence (so-called ``holography'')
\cite{M,W} between quantum field theories on 1+$s$-dimensional
anti-deSitter space-time $\ads$ (the ``bulk space'') and conformal 
quantum field theories on its conformal boundary $\cm$ which is a 
compactification of Min\-kow\-ski space $\RR^{1,s-1}$, has recently 
raised enthusiastic interest. If anti-deSitter space is considered as an
approximation to the space-time geometry near certain gravitational
horizons (extremal black holes), then the correspondence lends support
to the informal idea of reduction of degrees of freedom due to the
thermodynamic properties of black holes \cite{tH,Sk}. Thus, holography
is expected to give an important clue for the understanding of
quantum theory in strong gravitational fields and, ultimately, of
quantum gravity.  

While the original conjecture \cite{M}
was based on ``stringy'' pictures, it was soon formulated \cite{W} in
terms of (Euclidean) conventional quantum field theory, and a specific
relation between generating functionals was conjectured. These
conjectures have since been exposed with success to many structural
and group theoretical tests, yet a rigorous proof has not been given. 

The problem is, of course, that the ``holographic'' transition from 
anti-deSitter space to its boundary and back, is by no means a 
point transformation, thus preventing a simple (pointwise) operator 
identification between bulk fields and boundary fields. In
the present note, we show that in contrast, an identification between
the {\em algebras generated by}\/ the respective local bulk and boundary 
fields is indeed possible in a very transparent manner. These
algebraic data are completely sufficient to reconstruct the
respective theories. 

We want to remind the reader of the point of view due to Haag and Kastler
(see \cite{H} for a standard textbook reference) which emphasizes
that, while any choice of particular fields in a quantum field theory
may be a matter of convenience without affecting the physical content
of the theory (comparable to the choice of coordinates in geometry), the
algebras they generate and their algebraic interrelations, notably
causal commutativity, supply all the relevant physical information in
an invariant manner. The interested reader will find in \cite{B} a
review of the (far from obvious, indeed) 
equivalence between quantum field theory in terms of fields and
quantum field theory in terms of algebras, notably on the strategies
available to extract physically relevant information, such as
the particle spectrum, superselection charges, and scattering
amplitudes, from the net of algebras without knowing the fields. 

It is crucial in the algebraic approach, however, to keep track of the
{\em localization} of the 
observables. Indeed, the physical interpretation of a theory is coded
in the structure of a ``causal net'' of algebras \cite{H} which means
the specification of the sets of observables $B(X)$
which are localized in any given space-time region $X$.\footnote{The
  assignment $X \mapsto B(X)$ is a ``net'' in the mathematical sense:
  a generalized sequence with a partially ordered index set (namely
  the set of regions $X$).}     

The assignment $X \mapsto B(X)$ is subject
to the conditions of isotony (an observable localized in a region $X$
is localized in any larger region $Y \supset X$, thus $B(Y) \supset
B(X)$), causal commutativity (two observables localized
at space-like distance commute with each other), and covariance (the
Poincar\'e transform of an observable localized in $X$ is localized in
the transformed region $gX$; in the context at hand replace
``Poincar\'e'' by ``anti-deSitter''). Each $B(X)$ should in fact be an
algebra of operators (with the observables its selfadjoint elements),
and to have sufficient control of limits and convergence in order to compute
physical quantities of interest, it is convenient to let $B(X)$ be von Neumann
algebras.\footnote{A von Neumann algebra is an algebra of bounded
  operators on a Hilbert space which is closed in the weak topology of
  matrix elements. E.g., if $\phi$ is a hermitean field and $\phi(f)$
  a field operator smeared over a region $X$ containing the support of
  $f$, then operators like $\exp i\phi(f)$ belong to $B(X)$.} 
  
For most purposes it is convenient to consider as typical compact 
regions the ``double-cones'', that is, intersections of a future 
directed and a past directed light-cone, and to think of point-like
localization in terms of very small double-cones. On the other hand, 
certain aspects of the theory are better captured by ``wedge'' regions
which extend to space-like infinity. A space-like wedge (for short: wedge) in
Minkowski space is a region of the form $\{x: x_1>\vert x_0\vert\}$,
or any Poincar\'e transform thereof. The corresponding regions in
anti-deSitter space turn out to be intersections of $\ads$ with
suitable flat space wedges in the ambient space $\RR^{2,s}$, see below.
In conformally covariant theories there is no distinction between 
double-cones and wedges since conformal transformations map the former
onto the latter.

It will become apparent in the sequel that to understand the issue of
``holography'', the algebraic framework proves to be most appropriate. 

The basis for the holography conjectures is, of course, the
coincidence between the anti-deSitter group $SO_0(2,s)$ and the
conformal group $SO_0(2,s)$. ($SO_0(n,m)$ is the identity component of the
  group $SO(n,m)$, that is the proper orthochronous subgroup
  distinguished by the invariant condition that the determinants of
  the time-like $n \times n$ and of the space-like $m \times m$
  sub-matrices are both positive.) The former group acts on 
$\ads$ (as a ``deformation'' from the flat space Poincar\'e group
in 1+$s$ dimensions, $SO_0(1,s) \ltimes \RR^{1,s}$), 
and the latter group acts on the conformal boundary $\cm$ of $\ads$ 
(as an extension of the Poincar\'e group in 1+($s-1$) dimensions, 
$SO_0(1,s-1) \ltimes \RR^{1,s-1}$) by restriction of the former
group action on the bulk. The representation theoretical aspect of
this coincidence has been elaborated (in Euclidean metric) in \cite{D}. 

In terms of covariant nets of algebras of local observables (``local
algebras''), it is thus sufficient to identify one suitable algebra in
anti-deSitter space with another suitable algebra in the conformal
boundary space, and then to let $SO_0(2,s)$ act to provide the remaining
identifications. As any double-cone region 
in conformal space determines a subgroup of the conformal group $SO_0(2,s)$ 
which preserves this double-cone, it is natural to identify its algebra 
with the algebra of a region in anti-deSitter space which is preserved 
by the same subgroup of the anti-deSitter group $SO_0(2,s)$. It turns out 
that this region is a space-like wedge region 
which intersects the boundary in the given double-cone. 

For a typical bulk observable localized in a wedge region, the reader
is invited to think of a field operator for a Mandelstam string which
stretches to space-like infinity. Its holographic
localization on the boundary has finite size, but it becomes sharper
and sharper as the string is ``pulled to infinity''. We shall see that
one may be forced to take into consideration theories which possess only
wedge-localized, but no compactly localized observables. 

Our main algebraic result rests on the following geometric
Lemma:\footnote{%
%  The symbol $p$ stands for the projection from the 
%  hypersurface $H_{1,s}=\{x\in\RR^{2,s}:x^2=R^2\}$ onto its quotient $\ads$ by
%  the identification of points $x$ and $-x$. 
For details, see Sect.\ 2. We denote double-cones in
  the boundary by the symbol $I$, because {\sl (i)} we prefer to reserve the
  ``standard'' symbol $O$ for double-cones in the bulk space, and because 
  {\sl (ii)} in 1+1 dimensions the ``double-cones'' on the
  boundary are in fact open intervals on the circle $S^1$. } 

\vskip1mm{\bf Lemma: \sl Between the set of space-like wedge regions in 
  anti-deSitter space, $\pw \subset \ads$, and the set of
  double-cones in its conformal boundary space, $I \subset \cm$,
  there is a canonical bijection $\alpha: \pw \mapsto I=\a(\pw)$
  preserving inclusions and causal complements, and intertwining the
  actions of the anti-deSitter group $SO_0(2,s)$ and of the conformal
  group $SO_0(2,s)$ 
  $$\a(g(\pw)) = \dot g(\a(\pw)), \qquad \a\inv(\dot g(I))=g(\a\inv(I))$$
  where $\dot g$ is the restriction of the action of $g$ to the boundary. 
  The double-cone $I=\a(\pw)$ associated with a wedge $\pw$ is the
  intersection of $\pw$ with the boundary. }\vskip1mm

Given the Lemma, the main algebraic result states that bulk
observables localized in wedge regions are identified with 
boundary observables localized in double-cone regions:

\vskip1mm{\bf Corollary 1: \sl The identification of local observables
  $$ B(\pw) := A(\a(\pw)), \qquad A(I) := B(\a\inv(I)) $$
  gives rise to a 1:1 correspondence between isotonous causal conformally
  covariant nets of algebras $I \mapsto A(I)$ on $\cm$ and
  isotonous causal anti-deSitter covariant nets of algebras $\pw
  \mapsto B(\pw)$ on $\ads$.} \vskip1mm

An observable localized in a double-cone $O$ in anti-deSitter space is
localized in any wedge containing $O$, hence the local algebra $B(O)$
should be contained in all $B(\pw)$, $\pw \supset O$. We shall define
$B(O)$ as the intersection of all these wedge algebras.  
These intersections do no longer correspond to simple geometric
regions in $\cm$ (so points in the bulk have a complicated
geometry in the boundary), as will be discussed in more detail in 1+1
dimensions below.  

The following result also identifies states and representations of the
corresponding theories:

\vskip1mm{\bf Corollary 2: \sl Under the identification of Corollary 1, 
  a vacuum state on the net $A$ corresponds to a vacuum state on the net
  $B$. Positive-energy representations of the net $A$ correspond to
  positive-energy representations of the net $B$. The net $A$
  satisfies essential Haag duality if and only if the net $B$ does.
  The modular group and modular conjugation (in the sense of
  Tomita-Takesaki) of a wedge algebra $B(\pw)$ in a vacuum state 
  act geometrically (by a subgroup of $SO_0(2,s)$ which preserves
  $\pw$ and by a CPT reflection, respectively) if and
  only if the same holds for the double-cone algebras $A(I)$.  }\vskip1mm
% coincides with the modular
%  group of the corresponding double-cone algebra $A(I)$ in the
%  corresponding vacuum state, which is a one-parameter 
%  subgroup of $SO_0(2,s)$ preserving $\pw$ and $I$, respectively.
% The modular conjugation of a
%  wedge algebra $B(\pw)$ in a vacuum state is a CPT-type transformation
%  which maps $\pw$ onto its causal complement.

Essential Haag duality means that the algebras associated with
causally complementary wedges not only commute as required by locality, but
either algebra is in fact the {\em maximal} algebra commuting with the other
one. 

The last statement in the Corollary refers to the modular theory of
von Neumann algebras which states that every (normal and cyclic) state
on a von Neumann algebra is a thermal equilibrium state with respect
to a unique adapted ``time'' evolution (one-parameter group of
automorphisms = modular group) of the algebra. In quantum field
theories in Minkowski space, whose local algebras are generated by
smeared Wightman fields, the modular groups have been computed for
wedge algebras in the vacuum state \cite{BW} and were found to
coincide with the boost subgroup of the Lorentz group which preserves
the wedge (geometric action). In conformal theories, mapping wedges
onto double-cones by suitable conformal transformations, the same
result also applies to double-cones \cite{HL}. This result is an
algebraic explanation of the Unruh effect according to which a
uniformly accelerated observer attributes a temperature to the vacuum
state, and provides also an explanation of Hawking radiation if the
wedge region is replaced by the space-time region outside the horizon
of a Schwarzschild black hole \cite{S}.   

The modular theory also provides a ``modular conjugation'' which maps the
algebra onto its commutant. 
For Minkowski space Wightman field theories in the vacuum state as
before, the modular conjugation of a wedge algebra turns out to act 
geometrically as a CPT-type reflection (CPT up to a rotation) 
along the ``ridge'' of the wedge which maps the wedge onto its causal
complement. This entails essential duality for Minkowski space \cite{BW} as
well as conformally covariant \cite{HL} Wightman theories.

The statement in Corollary 2 on the modular group thus implies that, 
if the boundary theory is a Wightman
theory, then the boundary and the bulk theory both satisfy essential
Haag duality, and also in anti-deSitter space a vacuum state of 
$B$ in restriction to a
wedge algebra $B(\pw)$ is a thermal equilibrium state with respect to
the associated one-parameter boost subgroup of the anti-deSitter group
which preserves $\pw$, i.e., the Unruh effect takes place for a
uniformly accelerated observer. Furthermore, the CPT theorem holds for
the theory on anti-deSitter space. On the other hand, essential Haag
duality and geometric modular action for quantum field
theories on $\ads$ were established under much more general assumptions
\cite{BFS},
implying the same properties for the associated boundary theory even
when it is not a Wightman theory (see below).

We emphasize that the Hamiltonians $\frac 1R M_{0,d}$ on $\ads$
and $P^0$ on $\cm$ are {\em not}\/ identified under the
identification of the anti-deSitter group and the conformal group.
%(thus giving rise to different concepts of entropy, and different
%counting of degrees of freedom). 
Instead, $M_{0,d}$ is (in suitable
coordinates) identified with the combination $\frac12 (P^0+K^0)$ of
translations and special conformal transformations in the $0$-direction  
of $\cm$. This is different from the Euclidean picture \cite{W}
where the anti-deSitter Hamiltonian is identified with the dilatation
subgroup of the conformal group. In Lorentzian metric, the dilatations
correspond to a space-like ``translation'' subgroup of the
anti-deSitter group. This must have been expected since the generator
of dilatations does not have a one-sided spectrum as is required for
the real-time Hamiltonian. The subgroup generated by $\frac12(P^0+K^0)$ 
is well-known to be periodic and to satisfy the spectrum condition in
every positive-energy representation. (Periodicity in bulk time of
course implies a mass gap for the underlying bulk theory. This is not
in conflict with the boundary theory being massless since the
respective subgroups of time evolution cannot be identified.) 

Different Hamiltonians give rise to different counting of
degrees of freedom, since entropy is defined via the partition function.
Thus, the ``holographic'' reduction of degrees of freedom \cite{tH,Sk}
can be viewed as a consequence of the choice of the Hamiltonian: 
The anti-deSitter Hamiltonian $M_{0,d}=\frac12 (P^0+K^0)$ has discrete
spectrum and has a chance (at least in 1+1 dimensions) to yield a
finite partition function. One the other hand, the partition function
with respect to the boundary Hamiltonian $P^0$ exhibits the usual
infrared divergence due to infinite volume and continuous spectrum. 

A crucial aspect of the present analysis is the identification 
of compact regions in the boundary with wedge regions in the
bulk. With a little hindsight, this aspect is indeed also present in
the proposal for the identification of generating functionals
\cite{W}. While the latter is given in the Euclidean approach, 
it should refer in real time to a hyperbolic differential equation
with initial values given in a double-cone on the boundary which
determine its solution in a wedge region of bulk space.

We also show that in 1+1 dimensions there are sufficiently
many observables localized in arbitrarily small {\em compact}\/
regions in the bulk space to ensure that compactly localized
observables generate the wedge algebras. This property is crucial if
we want to think of local algebras as being generated by local fields:

\vskip1mm{\bf Proposition: \sl Assume that the boundary theory $A$ on $S^1$
  is weakly additive (i.e., $A(I)$ is generated by $A(J_n)$ whenever the
  interval $I$ is covered by a family of intervals $J_n$). If a wedge
  $\pw$ in $AdS_{1,1}$ is covered by a family of double-cones 
  $O_n \subset \pw$, then the algebra $B(\pw)$ is generated by the
  observables localized in $O_n$:}  
  $$B(\pw)=\bigvee_n B(O_n).$$ \vskip1mm

In order to establish this result, we explicitly determine the observables
localized in a double-cone region on $AdS_{1,1}$. Their algebra $B(O)$
turns out to be non-trivial: it is the intersection of two interval
algebras $A(I_i)$ on the boundary $S^1$ where the intersection of the two
intervals $I_i$ is a union of two disjoint intervals $J_i$. $B(O)$
contains therefore at least $A(J_1)$ and $A(J_2)$. In fact, it is even
larger than that, containing also observables corresponding to a
``charge transport'' \cite{H}, that is, operators which annihilate a
superselection charge in $J_1$ and create the same charge in
$J_2$. The inclusion $A(I_1)\vee A(J_2)\subset B(O)$ therefore
carries (complete) algebraic information about the superselection
structure of the chiral boundary theory \cite{KLM}. 

As the double-cone on $AdS_{1,1}$ shrinks, the size of the intervals
$J_i$ also shrinks but not their distance, so points in 1+1-dimensional  
anti-deSitter space are related to pairs of points in conformal space. 
But we see that sharply localized bulk observables involve boundary 
observables localized in large intervals: the above charge transporters. 
This result provides an algebraic interpretation of the obstruction
against a point transformation between bulk and boundary. 

The issue of compactly localized observables in anti-deSitter space 
is more complicated in more than two dimensions, and deserves a
separate careful analysis. Some preliminary results will be presented
in Section 2.3. They show that if the bulk theory possesses observables
localized in double-cones, then the corresponding boundary theory
violates an additivity property which is characteristic for 
Wightman field theories, while its violation is expected for
non-abelian gauge theories due to the presence of gauge-invariant
Wilson loop operators. Conversely, if the boundary theory satisfies
this additivity property, then the observables of the corresponding
bulk theory are always attached to infinity, as in topological
(Chern-Simons) theories.    

Let us point out that the conjectures in \cite{M,W} suggest a much
more ambitious interpretation, namely that the correspondence pertains
to bulk theories {\em involving quantum gravity}, while the
anti-deSitter space and its boundary are understood in some asymptotic
(semi-classical) sense. Indeed, the algebraic approach is no more
able to describe proper quantum gravity as any other mathematically
unambiguous framework up to now. Most arguments given in the
literature in favour of the conjectures refer to gravity as
perturbative gravity on a background space-time. Likewise, our present
results concern the semi-classical version of the conjectures,
treating gravity like any other quantum field theory as a theory of
observables on a classical background geometry. In fact, the presence
or absence of gravity in the bulk theory plays no particular 
role. This is only apparently in conflict with the original arguments 
for a holographic reduction of degrees of freedom of a bulk theory in
the vicinity of a gravitational horizon \cite{tH,Sk} in which gravity
is essential. Namely, our statement can be interpreted in the sense
that {\em once the anti-deSitter geometry is given} for whatever reason
(e.g., the presence of a gravitational horizon), then it can support 
only the degrees of freedom of a boundary theory. 

\section{Identification of observables}

We denote by $H_{1,s}$ the $d$=1+$s$-dimensional hypersurface defined through 
its embedding into ambient $\RR^{2,s}$,
$$x_0^2-x_1^2-\dots-x_s^2+x_d^2=R^2$$
with Lorentzian metric induced from the 2+$s$-dimensional metric
$$ds^2=dx_0^2-dx_1^2-\dots-dx_s^2+dx_d^2.$$
Its group of isometries is the Lorentz group $O(2,s)$ of the ambient space 
in which the reflection $x\mapsto -x$ is central. Anti-deSitter space
is the quotient manifold $\ads=H_{1,s}/\ZZ_2$ (with the same Lorentzian 
metric locally). We denote by $p$ the projection $H_{1,s} \to \ads$.

Two open regions in anti-deSitter space are called ``causally 
disjoint'' if none of their points can be connected by a time-like 
geodesic. The largest open region causally disjoint from a given region is
called the causal complement. In a causal quantum field theory on the
quotient space $\ads$, observables and hence algebras associated with
causally disjoint regions commute with each other.   

The reader should be worried about this definition, since causal
independence of observables should be linked to causal connectedness
by time-like {\em curves}\/ rather than geodesics. But on anti-deSitter 
space, any two points can be connected by a time-like curve, so they
are indeed causally connected, and the requirement that causally 
{\em disconnected}\/ observables commute is empty. Yet, as our
Corollary 1 shows, if the boundary theory is causal, then the
associated bulk theory is indeed causal in the present (geodesic)
sense. We refer also to \cite{BFS} where it is shown that vacuum
expectation values of commutators of observables with causally
disjoint localization have to vanish whenever the vacuum state has 
reasonable properties (invariance and thermodynamic passivity), but
without any a priori assumptions on causal commutation relations
(neither in bulk nor on the boundary). 

Thus in the theories on anti-deSitter space we consider in this paper,
observables localized in causally disjoint but causally connected
regions commute; see \cite{BFS} for a discussion of the ensuing 
physical constraints on the nature of interactions on anti-deSitter space.

The causal structure of $\ads$ is determined by its metric modulo 
conformal transformations which preserve angles and geodesics. As a causal 
manifold, $\ads$ has a boundary (the ``asymptotic directions'' of 
geodesics). The boundary inherits the causal structure of the bulk space 
$\ads$, and the anti-deSitter group $SO_0(2,s)$ acts on this space. 
It is well known that this boundary is a compactification $\cm
=(S^1\times S^{s-1})/\ZZ_2$ 
of Minkowski space $\RR^{1,s-1}$, and $SO_0(2,s)$ acts on it like the
conformal group. 

The notions of causal disjoint and causal complements on $\cm$
coincide, up to conformal transformations, with those on Minkowski space 
\cite{LM}. In $d=$1+1 dimensions, $s=1$, the conformal space is $S^1$,
and the causal complement of an interval $I$ is $I^c = S^1 \backslash
\overline I$. 

Both anti-deSitter space and its causal boundary have a ``global
time-arrow'', that is, the distinction between the future and past
light-cone in the tangent spaces (which are ordinary Minkowski spaces) 
at each point $x$ can be globally chosen continuous in $x$ (and 
consistent with the reflection $x \mapsto -x$). The time orientation
on the bulk space induces the time orientation on the boundary. The
time arrow is crucial in order to distinguish representations of
positive energy.

\vskip2mm {\bf 2.1 Proof of the Lemma}

Any ordered pair of light-like vectors $(e,f)$ in the ambient
space $\RR^{2,s}$ such that $e \ccdot f < 0$ defines an open subspace of
the hypersurface $H_{1,s}$ given by 
$$\wt(e,f) = \{x \in \RR^{2,s}: x^2 = R^2, e \ccdot x > 0, f \ccdot x > 0 \}.$$
This space has two connected components. Namely, the tangent vector at
each point $x \in \wt(e,f)$ under the boost in the $e$-$f$-plane,
$\delta_{e,f}\, x = (f \ccdot x)e - (e \ccdot x) f$, is either a
future or a past directed time-like vector, since $(\delta_{e,f}\, x)^2 
= -2(e \ccdot f) (e \ccdot x) (f \ccdot x)>0$. We denote by $\wt_+(e,f)$
and $\wt_-(e,f)$ the connected components of $\wt(e,f)$ in which
$\delta_{e,f}\, x$ is future and past directed, respectively. 
By this definition, $\wt_+(f,e) = \wt_-(e,f)$, and $\wt_+(-e,-f) = -\wt_+(e,f)$.   

The wedge regions in the hypersurface $H_{1,s}$ are the regions
$\wt_\pm(e,f)$ as specified. The wedge regions in anti-deSitter space
are their quotients $\pw_\pm(e,f)=p\wt_\pm(e,f)$. One has
$\pw_+(e,f)=\pw_-(f,e)=\pw_+(-e,-f)$, and $\pw_+(e,f)$ and
$\pw_-(e,f)$ are each other's causal complements. 
For an illustration in 1+1 dimensions, cf.\ Figure 1.

\hskip10mm\epsfig{file=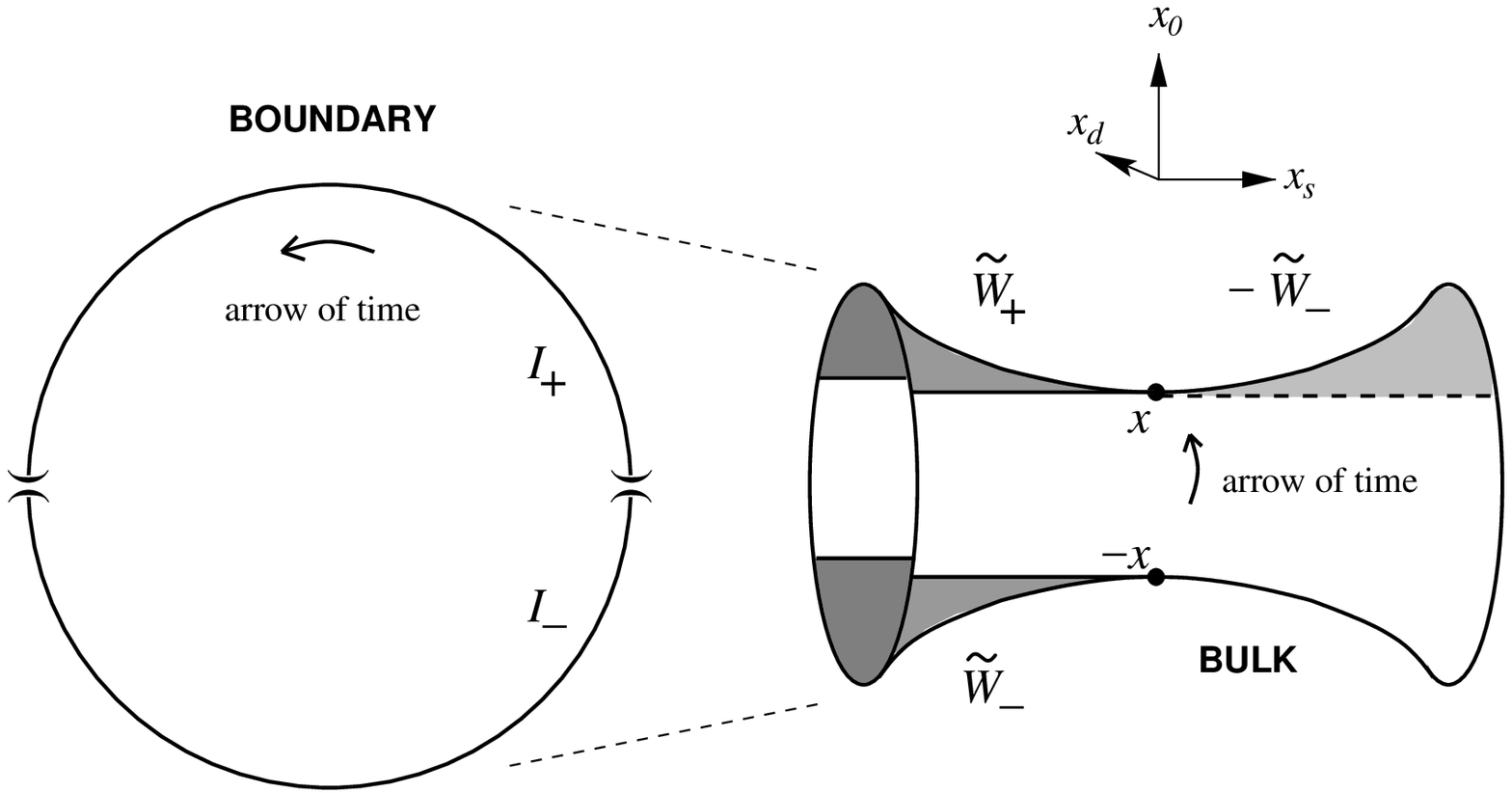,width=14cm}

{\bf Figure 1. \sl Wedge regions $\wt_+(e,f)$ and $\wt_-(e,f)$ in 1+1
  dimensions, and their intersections with the boundary. The
  light-like vectors $e$ and $f$ are tangent to $-\wt_-$ in its apex
  $x$. In anti-deSitter space, $\wt_-$ is identified with $-\wt_-$, and 
  $\pw_\pm=p\wt_\pm$ are causal complements of each other.} \vskip1mm

We claim that the projected wedges $\pw_\pm(e,f)$ intersect the
boundary of $\ads$ in regions $I_\pm(e,f)$ which  are
double-cones of Minkowski space $\RR^{1,s-1}$ or images thereof under
some conformal transformation. Note that any two double-cones in
$\RR^{1,s-1}$ are connected by a conformal transformation, and among
their conformal transforms are also the past and future light-cones
and space-like wedges in $\RR^{1,s-1}$.  

We claim also that the causal complement $\pw_-(e,f)$ of
the wedge $\pw_+(e,f)$ intersects the boundary in the causal complement
$I_-(e,f)=I_+(e,f)^c$ of  $I_+(e,f)$. 

It would be sufficient to compute the intersections of any single pair of
wedges $\wt_\pm(e,f)$ with the boundary and see that it is a pair of 
causally complementary conformal double-cones in $\cm$, since
the claim then follows for any other pair of wedges by covariance. For
illustrative reason we shall compute two such examples. 

We fix the ``arrow of time'' by declaring the tangent vector of the
rotation in the 0-$d$-plane, $\delta_t\, x=(-x_d,0,\dots,0,x_0)$, to be
future directed.

In stereographic coordinates $(y_0,\vec y,x_-)$ of the hypersurface 
$x^2=R^2$, where $x_-=x_d-x_s$ and $(y_0,\vec y)=(x_0,\vec x)/x_-$,
$\vec x=(x_1,\dots,x_{s-1})$, the boundary 
is given by $\vert x_- \vert = \infty$. Thus, in the limit of infinite
$x_-$ one obtains a chart $y_\mu=(y_0,\vec y)$ of $\cm$. The
induced conformal structure is that of Minkowski space,
$dy^2=f(y)^2(dy_0^2-d\vec y^2)$.

Our first example is the one underlying Figure 1%for 1+1 dimensions
: we choose $e_\mu=(0,\dots,0,1,1)$ and
$f_\mu=(0,\dots,0,1,-1)$. %\footnote{This is the choice underlying
%  Figure 1 for 1+1 dimensions.}
The conditions for $x \in \wt(e,f)$ read $x_d-x_s>0$ and $x_d+x_s<0$, implying
$x_d^2-x_s^2<0$ and hence $x_0^2-\vec x^2 > R^2$. The tangent vector
$\delta_{e,f}\, x$ has $d$-component $\delta_{e,f}\, x_d=-2x_s>0$. Hence, it 
is future directed if $x_0>0$, and past directed if $x_0<0$:
$$ \wt_+(e,f)=\{x: x^2=R^2, x_s<-\vert x_d \vert, x_0>0\}.$$
After dividing $(x_0,\vec x)$ by $x_- = x_d-x_s\nearrow \infty$,
we obtain the boundary region
$$I_+(e,f)=\{y=(y_0,\vec y): y_0^2-\vec y^2>0,y_0>0\},$$
that is, the future light-cone in Minkowski space; similarly, $I_-(e,f)$ 
is the past light-cone, and $I_\pm(e,f)$ are each other's causal complements 
in $\cm$.  

Next, we choose $e_\mu=(1,1,0,\dots,0)$ and $f_\mu=(-1,1,0,\dots,0)$.
The conditions for $x \in \wt(e,f)$ read $x_1 < -\vert x_0 \vert$, implying
$x_0^2-\vec x^2 < 0$ and hence $x_d^2-x_s^2 > R^2 >0$. The tangent vector
$\delta_{e,f}\, x$ has $0$-component $\delta_{e,f}\, x_d=-2x_1>0$. Hence, it 
is future directed if $x_d<0$ hence $x_d-x_s<0$, and past directed if 
$x_d>0$ hence $x_d-x_s>0$:
$$ \wt_+(e,f)=\{x: x^2=R^2, x_1<-\vert x_0 \vert, x_-<0\}.$$
After dividing $(x_0,\vec x)$ by $x_-=x_d-x_s\searrow -\infty$, we
obtain the boundary region 
$$I_+(e,f)=\{y=(y_0,\vec y): y_1 > \vert y_0 \vert\},$$
that is, a space-like wedge region in Minkowski space; similarly, 
$I_-(e,f)$ is the opposite wedge $y_1 < -\vert y_0 \vert$, which is
again the causal complement of $I_+(e,f)$ in $\cm$.   

Both light-cones and wedge regions in Minkowski space are well known to 
be conformal transforms of double-cones, and hence they are
double-cones on $\cm$. The pairs of regions computed above are
indeed causally complementary pairs. 

We now consider the map $\alpha: \pw_+(e,f) \mapsto I_+(e,f)$.
Since the action of the conformal group on the boundary is induced by 
the action of the anti-deSitter group on the bulk, we see that
$\wt_\pm(ge,gf)=g(\wt_\pm(e,f))$ and $I_\pm(ge,gf)=\dot g(I_\pm(e,f))$, hence
$\alpha$ intertwines the actions of the anti-deSitter and the conformal 
group. It is clear that $\alpha$ preserves inclusions, and we have seen 
that it preserves causal complements for one, and hence for all wedges.
Since $SO_0(2,s)$ acts transitively on the set of double-cones of
$\cm$, the  
map $\alpha$ is surjective. Finally, since $\pw_+(e,f)$ and $I_+(e,f)$ have 
the same stabilizer subgroup of $SO_0(2,s)$, it is also injective. 

This completes the proof of the Lemma. \qed

\vskip2mm {\bf 2.2 Proof of the Corollaries}

We identify wedge algebras on $\ads$ and double-cone algebras
on $\cm$ by 
$$ B(\pw_\pm(e,f)) = A(I_\pm(e,f)),$$
that is, $B(\pw)=A(\a(\pw))$. 
The Lemma implies that if $A$ is given as an isotonous, causal and
conformally covariant net of algebras on $\cm$, then $B(\pw)$
defined by this identification constitute an isotonous, causal and
anti-deSitter covariant net of algebras on $\ads$, and vice
versa. Namely, the identification is just a relabelling of the index
set of the net which preserves inclusions  and causal complements and
intertwines the action of $SO_0(2,s)$. 
Thus we have established Corollary 1. \qed

As for Corollary 2, we note that, as the algebras
are identified, states and representations of the nets $A$ and $B$ are
also identified. 

Since the identification intertwines the action of the anti-dSitter group
and of the conformal group, an anti-deSitter invariant state on $B$
corresponds to a conformally invariant state on $A$. The generator of
time translations in the anti-deSitter group corresponds to the
generator $\frac12 (P^0+K^0)$ in the conformal group which is known to
be positive if and only if $P^0$ is positive (note that $K^0$ is
conformally conjugate to $P^0$). Hence the conditions for positivity of
the respective generators of time-translations are equivalent.

By the identification of states and algebras, also the modular groups
are identified. The modular group and modular conjugation for
double-cone algebras in a vacuum state of conformally covariant
quantum field theories are conformally conjugate to the
modular group and modular conjugation of a Minkowski space 
wedge algebra, which in turn are given by the Lorentz boosts in the
wedge direction and the reflection along the ridge of the wedge
\cite{BW,HL}. It follows that the modular group for a wedge algebra on 
anti-deSitter space is given by the corresponding subgroup of the
anti-deSitter group which preserves the wedge (for a wedge $\pw_+(e,f)$,
this is the subgroup of boosts in the $e$-$f$-plane), and the
modular conjugation is a CPT transformation which maps $\pw_+$ onto $\pw_-$.

These remarks suffice to complete the proof of Corollary 2. \qed

Let us mention that the correspondence given in Corollary 1 holds also 
for ``weakly local'' nets both on the bulk and on the boundary. In a weakly 
local net, the vacuum expectation value of the commutator of two 
causally disjoint observables vanishes, but not necessarily the commutator 
itself. Weak locality for quantum field theories on anti-deSitter space 
follows \cite{BFS} from very conservative assumptions on the vacuum
state without any commutation relations assumed. Thus, also the boundary
theory will always be weakly local. 

\vskip2mm {\bf 2.3 Compact localization in anti-deSitter space}

Let us first note that as the ridge of a wedge is shifted into the
interior of the wedge, the double-cone on the boundary shrinks. Thus,
sharply localized boundary observables correspond to bulk observables
at space-like infinity \cite{BBMS}. We now show that sharply localized bulk
observables do not correspond to a simple geometry on the boundary,
but must be determined algebraically. 

An observable localized in a double-cone $O$ of anti-deSitter space must
be contained in every wedge algebra $B(\pw)$ such that $O \subset \pw$. 
The algebra $B(O)$ is thus at most the intersection of all $B(\pw)$
such that $O \subset \pw$. We may define it as this intersection,
thereby ensuring isotony, causal commutativity and covariance for the
net of double-cone algebras in an obvious manner.  

Double-cone algebras on anti-deSitter space are thus delicate 
intersections of algebras of double-cones and their conformal images on 
the boundary, and might turn out trivial. In 1+1 dimensions, the
geometry is particularly simple since a double-cone is an intersection
of only two wedges. We show that the corresponding intersection of
algebras is non-trivial, and shall turn to $d>1+1$ below. 

Let us write (in 1+1 dimensions) the relation
$$B(O)=B(\pw_1) \cap B(\pw_2) \equiv A(I_1) \cap A(I_2)
\qquad\hbox{whenever}\qquad O=\pw_1 \cap \pw_2,$$
where $\pw_i$ are any pair of wedge regions in $AdS_{1,1}$ and $I_i=\a(\pw_i)$
their intersections with the boundary, that is, open intervals on $S^1$.

The intersection $\pw_1 \cap \pw_2$ might not be a double-cone. It might
be empty, or it might be another wedge region. Before discussing the
above relation as a definition for the double-cone algebra $B(O)$ if 
$O=\pw_1 \cap \pw_2$ is a double-cone, we shall first convince ourselves
that it is consistent also in these other cases. 

If $\pw_1$ contains $\pw_2$, or vice versa, then $O$ equals the larger
wedge, and the relation holds by isotony. 
If $\pw_1$ and $\pw_2$ are disjoint, then the intersections with the
boundary are also disjoint, and $B(\emptyset) = A(I_1) \cap A(I_2)$ is
trivial if the boundary net $A$ on $S^1$ is irreducible (that is, 
disjoint intervals have no nontrivial observables in common). 

Next, it might happen that $\pw_1$ and $\pw_2$ have a nontrivial
intersection without the apex of one wedge lying inside the other
wedge. In this case, the intersection is again a wedge, say $\pw_3$.
Namely, any wedge in $AdS_{1,1}$ is of the form $\pw_+(e,f)$ where $e$
and $f$ are a future and a past directed light-like tangent vector in 
the apex $x$ (the unique point in $AdS_{1,1}$ solving $e \ccdot x = f
\ccdot x = 0$). The condition $e \ccdot f < 0$ implies that both tangent
vectors point in the same (positive or negative) 1-direction. The
wedge itself is the surface between the two light-rays emanating 
from $x$ in the directions  $-e$ and $-f$ (cf.\ Figure 1).
The present situation arises if the future directed light-ray of
$\pw_1$ intersects the past directed light-ray of $\pw_2$ (or vice
versa) in a point
$x_3$ without the other pair of light-rays intersecting each other. 
The intersection of the two wedges is the surface between the two
intersecting light-rays travelling on from the point $x_3$, which is
another wedge region $\pw_3$ with apex $x_3$. It follows that the
intersection of the intersections $I_i$ of $\pw_i$ with the boundary equals 
the intersection $I_3$ of $\pw_3$ with the boundary. Hence consistency 
of the above relation is guaranteed by $A(I_1) \cap A(I_2)=A(I_3)$ 
where $I_1$ and $I_2$ are two intervals on $S^1$ whose intersection 
$I_3$ is again an interval.

Now we come to the case that $\pw_1 \cap \pw_2$ is a double-cone $O$ in the
proper sense. This is the case if the closure of the causal complement of 
$\pw_1$ is contained in $\pw_2$. It follows that the closure of the
causal complement of $I_1$ is contained in $I_2$, hence the
intersection of $I_1$ and $I_2$ is the union of two disjoint intervals
$J_1$ and $J_2$. The latter are the two light-like geodesic ``shadows'', 
cast by $O$ onto the boundary.

Thus, the observables localized in a double-cone in anti-deSitter space 
$AdS_{1,1}$ are given by the intersection of two interval algebras $A(I_i)$ 
on the boundary for intervals $I_i$ with disconnected intersections 
(or equivalently, by essential duality, the joint commutant of two interval 
algebras for disjoint intervals). Such algebras have received much 
attention in the literature \cite{SW,X,KLM}, notably within the 
context of superselection sectors. Namely, if $I_1 \cap I_2 = J_1 \cup
J_2$ consists of two disjoint intervals, then the intersection of
algebras $A(I_1) \cap A(I_2)$ is larger than the algebra 
$A(J_1) \vee A(J_2)$. 
The excess can be attributed to the existence of superselection sectors 
\cite{KLM}, the extra operators being intertwiners which transport a 
superselection charge from one of the intervals $J_i$ to the other. 

We conclude that (certain) compactly localized observables on
anti-deSitter space are strongly delocalized observables
(charge transporters) of the boundary theory. Yet there is no
obstruction against both theories being Wightman theories generated by
local Wightman fields, as the following simple example shows.

In suitable coordinates $x_\mu=R\cdot(\cos t,\cos x,\sin t)/\sin x$,  
the bulk is the strip $(t,x)\in \RR\times (0,\pi)$ with points
$(t,x)\sim (t+\pi,\pi-x)$ identified, while the boundary are the points
$(0,u)$, $u\in\RR\;\hbox{mod}\;2\pi$. The metric is a multiple of
$dt^2-dx^2$, thus the light rays emanating from the bulk point $(t,x)$
hit the boundary at the points $u_\pm=t\pm x\;\hbox{mod}\;2\pi$. We see
that, as the double-cone $O$ shrinks to a point $(t,x)$ in bulk, the
two intervals $J_i$ on the boundary also shrink to points (namely $u_\pm$) 
while their distance remains finite. 

Now, we consider the abelian current field $j(u)$ on the boundary, and
determine the associated fields on anti-deSitter space. First, for
$(t,x)$ in the strip, both $j(t\pm x)$ are localized at $(t,x)$ and
give rise to a conserved vector current $j^\mu$ with components
$j^0(t,x)=j(t+x)+j(t-x)$, $j^1(t,x)=-j(t+x)+j(t-x)$. Furthermore, the
fields $\phi_\alpha(t,x)=\exp i\alpha\int_{t-x}^{t+x}j(u)du$ (suitably
regularized, of course), $\alpha\in \RR$, are also localized at $(t,x)$. 
Namely, since the charge operator $\int_{S^1}j(u)du$ is a number $q$
in each irreducible representation, $\phi_\alpha(t,x)$ may as well be
represented as $e^{i\alpha q} \exp -i\alpha\int_{t+x-2\pi}^{t-x} j(u)du$ 
and hence is localized in both complementary boundary intervals
$[t-x,t+x]$ and $[t+x-2\pi,t-x]$ which overlap in the points $u_+$ and
$u_-$, as required.  

Indeed, the fields $\phi_\alpha$ can be obtained from bounded Weyl
operators with finite localization as follows. $A(I)$ is generated by
boundary observables of the Weyl form $W(f)=\exp ij(f)$ where $f$ is a
smearing function on $S^1$ which is constant outside the interval
$I$. Adding a constant to $f$ is immaterial for the localization
since the commutation relations are given by the symplectic form 
$\int f'g\;du$. A Weyl operator $W(f)$ is localized in both intervals
$I_1$, $I_2$ (notation as before) if $f$ has constant values on both gaps
between $J_1$, $J_2$, but it is not a product of Weyl operators in
$J_1$ and in $J_2$ whenever these values are different. As a bulk
observable, $W(f)$ is localized in the double-cone $O=W_1\cap W_2$,
and operators of this form generate $B(O)$. Suitably regularized
limits of $W(f)$ yield the point-like local fields $\phi_\alpha(t,x)$.

For the more expert reader, we mention that our identification of 
double-cone algebras in bulk with two-interval algebras on the
boundary also shows how the notorious difficulty to compute the
modular group for two-interval algebras \cite{SW} is related to the
difficulty to compute the modular group of double-cone algebras in 
massive theories. (We discuss below that in a scaling limit the
massive anti-deSitter theory approaches a conformal flat space theory.
In this limit, the modular group can again be computed.) 

We now prove the Proposition of Sect.\ 1. It asserts that the
algebras $B(O_n)$ generate $B(\pw)$ whenever a family of double-cones
$O_n \subset \pw$ covers the wedge $\pw \subset AdS_{1,1}$. 

Each $B(O_n)$ is of the form $A(I_{n1}) \cap A(I_{n2})$ where $I_{n1}
\subset I = \alpha(\pw)$ and $I_{n1} \cap I_{n2} = J_{n1} \cup J_{n2}$ 
is a union of two disjoint intervals. By definition, the assertion is 
equivalent to
$$A(I) = \bigvee_n A(I_{n1}) \cap A(I_{n2}),$$
where the inclusion ``$\supset$'' holds since each $A(I_{n1})$ is contained 
in $A(I)$. On the other hand, the algebras on the right hand side are 
larger than $A(J_{n1}) \vee A(J_{n2})$. If $O_n$ cover the wedge $\pw$, 
then the intervals $J_{n1}$ and $J_{n2}$, as $n$ runs, cover the interval 
$I = \alpha(\pw)$. So the claim follows from weak additivity of the 
boundary theory. \qed

In $d\geq 2+1$ dimensions, the situation is drastically different.
Namely, if a family of small boundary double-cones $I_i$ covers the
space-like basis of a large double-cone $I$, and $\pw_i$
and $\pw$ denote the associated anti-deSitter wedge regions, then -- unlike in
1+1 dimensions -- $\pw$ will
contain a bulk double-cone $O$ which is space-like to all $\pw_i$.
Consequently, $B(O) \subset B(\pw)=A(I)$ must commute with the algebra
$\bigvee_i A(I_i)$ generated by all $B(\pw_i)=A(I_i)$. But in 
theories based on gauge-invariant Wightman fields (with the localization
of operators determined in terms of smearing functions), the latter
algebra coincides with $A(\bigcup_i I_i)$. This algebra in turn coincides with
$A(I)$ whenever the dynamics is generated by a Hamiltonian which is an
integral over a local density, because then the observables in a
neighbourhood
of the space-like basis of $I$ determine the observables in all of $I$.
Thus $B(O)$ must belong to the center of $A(I)$ which is commutative
(classical). Hence, a Wightman boundary theory is associated with a
bulk theory without compactly localized quantum observables.

Conversely, if there are double-cone localized bulk observables (e.g.,
if the bulk theory is itself described by a Wightman field \cite{F}), 
then the nontriviality of $B(O)$ requires $A(\bigcup_i I_i)=A(I)$ to be
strictly larger than $\bigvee_i A(I_i)$. This violation of
additivity seems to be characteristic of non-abelian gauge theories
where Wilson loop operators are not generated by point-like gauge
invariant fields (cf.\ also the discussion in \cite{ST}).

These issues certainly deserve a more detailed and careful
analysis. For the moment, we conclude that the holographic
correspondence necessarily relates, in more than 1+1 dimensions,
Wightman type boundary theories to bulk theories without
compactly localized observables (topological theories), in agreement
with a remark on Chern-Simons theories in \cite{W}, and, conversely,
bulk theories with point-like fields to boundary theories which
share properties of non-abelian gauge theories, in agreement with the
occurrence of Yang-Mills theory in \cite{M}.

\section{Speculations}

It is an interesting side-aspect of the last remark in the previous
section that the holographic correspondence in both
directions relates gauge theories to Wightman theories. It might
therefore provide a new constructive scheme giving access to gauge
theories.

If one is interested in quantum field theories on Minkowski 
rather than anti-deSitter space, one may consider the flat space 
limit in which the curvature radius $R$ of anti-deSitter space tends
to infinity, or equivalently consider a region of anti-deSitter
space which is much smaller than the curvature radius. The regime
$\vert x \vert << R$ asymptotically becomes flat Minkowski space, and
the anti-deSitter group contracts to the Poincar\'e group. Thus, 
one obtains a Minkowski space theory on $\RR^{1,s}$ from a conformal
theory on $\cm$ through a scaling limit \cite{B,BV} of the associated
theory on $\ads$. 

For $d=1+1$, this can be done quite explicitly. The double-cones
algebras $B(O)$ are certain extensions of the algebras $A(J_1) \vee
A(J_2)$, as discussed before. Now in the flat regime the intervals
$J_i$ become small of order $\vert O \vert/R$. Thus for a substantial
portion of Minkowski space, the relevant intervals $J_i$ are all
contained in a suitable but fixed pair of non-overlapping intervals
$K_i$. Let us now assume that the conformal net has the split property
(an algebraic property valid in any chiral quantum field theory for
which \hbox{Tr exp $-\beta L_0$} exists), which ensures that states
can be independently prepared on causally disjoint regions with a
finite distance \cite{B}. Then $A(K_1) \vee A(K_2)$ is unitarily
isomorphic to $A(K_1) \otimes A(K_2)$, and the isomorphism is
inherited by all its subalgebras $A(J_1) \vee A(J_2) \simeq A(J_1)
\otimes A(J_2)$. Under this isomorphism, the larger algebra $B(O)$ is
identified \cite{KLM} with the standard construction \cite{LR} of
1+1-dimensional conformal Minkowski space observables from a given
chiral conformal net (which corresponds to the diagonal modular
invariant and is sometimes quoted as the Longo-Rehren net): $B(O)
\simeq B_{\rm LR}(J_1 \times J_2)$ if $O$ corresponds to $I_1 \cap
I_2=J_1 \cup J_2 \subset K_1 \cup K_2$. The unitary isomorphism of
algebras, however, does not take the vacuum state on $B$ to the vacuum
state on the LR net. 

Thus, the flat space limit of the anti-deSitter space theory in 1+1 
dimensions associated with a given chiral conformal theory, is given by the 
LR net associated with that same chiral theory. Note that the LR net has 
1+1-dimensional conformal symmetry, but of course the anti-deSitter 
net is not conformally invariant due to the presence of the curvature
scale $R$. 

It would be interesting to get an analogous understanding of 
the flat space limit of the anti-deSitter space theory in higher
dimensions in terms of the associated conformal theory.  

One might speculate whether one can ``iterate holography'',
and use the flat space limit of the bulk theory on $\ads$ as a 
boundary input for a new bulk theory on $AdS_{1,s+1}$. Here, however,
a warning is in order. Namely, the limiting flat space theory on
Minkowski space $\RR^{1,s}$ will, like the LR net, in general not be 
extendible to the conformal compactification $C\!M_{1,s}$ but rather to a
covering thereof. One might therefore endeavour to extend the present
analysis to theories on covering spaces both of anti-deSitter space
and of its boundary. 

There is an independent and physically motivated reason to study
quantum field theories on a covering of anti-deSitter space. 
Namely, it has been observed (see above, \cite{BFS}) that the local
commutativity for causally disjoint but not causally disconnected
observables leads to severe constraints on the possible interactions
on anti-deSitter space proper. These constraints will disappear on the
universal covering space. 

This ``anti-deSitter causality paradox'' parallels very much the old 
``conformal causality paradox'' that causal commutativity on $C\!M_{1,s}$
proper excludes most conformal theories of interest; it was solved 
\cite{SS} by the recognition that conformal fields naturally live on a
covering space. Holography tells us that both problems are the
two sides of the same coin.

Extending the present analysis to covering spaces seems a dubious task
for $d=1+1$ since the boundary of the covering of two-dimensional
anti-deSitter space has two connected components. In higher
dimensions, however, we do not expect serious obstacles. 

\vskip5mm
{\large\bf Acknowledgment} 

Previous versions of this paper have been
improved in several respects on the basis of discussions with D. Buchholz,
R. Verch (who also made the Figure), B. Schroer, R. Helling and many
others, as well as questions raised by the referees. My thanks are due
to all of them. 

\vskip5mm
\small\addtolength{\baselineskip}{-1pt}

\end{document}